\documentclass[runningheads]{svmult}
\usepackage{makeidx}   % allows index generation
\usepackage{graphicx}  % standard LaTeX graphics tool
\usepackage{subeqnar}  % subnumbers individual equations
\usepackage{multicol}  % used for the two-column index
\usepackage{physprbb}  % modified textarea for proceedings,
\makeindex             % used for the subject index
\usepackage{amsmath}   % useful for coding complex math
\usepackage{amssymb}
\usepackage[latin1]{inputenc}
\begin{document}
\title*{Full Counting Statistics in Quantum Contacts} 

\titlerunning{Full Counting Statistics} 
\author{Wolfgang Belzig}  
\institute{Department of Physics and Astronomy, University of Basel,
  Klingelbergstr.82, CH-4056 Basel, Switzerland}

\maketitle

\begin{abstract}
  Full counting statistics is a fundamentally new concept in quantum
  transport. After a review of basic statistics theory, we introduce
  the powerful Green's function approach to full counting statistics.
  To illustrate the concept we consider a number of examples. For
  generic two-terminal contacts we show how counting statistics
  elucidates the common (and different) features of transport between
  normal and superconducting contacts. Finally, we demonstrate how
  correlations in multi-terminal structures are naturally included in
  the formalism.
\end{abstract}

\section{Introduction}
\label{sec:intro}

The probabilistic interpretation is a fundamental ingredient of
quantum mechanics. While the wave function determines the full quantum
state a system and its evolution in time, observable quantities are
related to hermitian operators. Expectation values of these operators
determine the average value of a large number of identical
measurements. However, an individual measurement yields in general a
different result.
%% An observable quantity is associated to a
%% hermitian operator. The expectation value of this operator in a given
%% state determines the average of a large number of measurements. The
%% outcome of an individual measurement, however, yields in general a
%% different value.  The probability to observe a certain value is
%% determined by a probability distribution.  E.~g., in standard wave
%% mechanics this fundamental quantity is the absolute square of the wave
%% function, which is a solution of the Schr\"odinger equation.
Applying this idea to a current measurement in a quantum conductor,
leads directly to the concept of \textit{full counting statistics}
(FCS): during a given time interval a certain number of charges will
pass the conductor.
%(determined e.g. by filtering
% Acquiring a current signal over some time (set e.g. by
%filtering) measures effectively the total charge that has passed. 
To predict the statistical properties of the number of transfered
charges we need a probability distribution. The theoretical goal is to
find this distribution.

\subsubsection{Overview}

In this article we give an introduction to the field of \textit{full
  counting statistics in mesoscopic electron transport}.  We will
concentrate on the powerful technique -- using Keldysh-Green's
functions -- which at the same time also is based on microscopic
theory. To accomplish this goal we will first review concepts of basic
statistics, which are relevant for counting statistics. In the next
section we address the microscopic derivation of FCS using
Keldysh-Green's functions. In the rest of the article we demonstrate
the use of counting statistics in a number of examples, like
two-terminal contacts with normal and superconducting leads, diffusive
metals and, finally, multi-terminal structures.  But first we review
briefly the development of the field.

\subsubsection{History}

Full counting statistics has its roots in quantum optics
\cite{mandelwolf}, where the number statistics of photons is used,
e.~g., to characterize coherence properties of photon sources.  The
major step to adopt the concept to mesoscopic electron transport has
been undertaken by Levitov and Lesovik \cite{levitov:93-fcs}.  Since
then the theory of FCS of charge transport in mesoscopic conductors
has advanced substantially, see Refs.~\cite{blanter,nazarov:03-book}.
In Ref.~\cite{levitov:93-fcs} it was shown that scattering between
uncorrelated Fermi leads with probability $T$ is described by a
binomial statistics $P(N)={ M \over N} T^N (1-T)^{M-N}$. Here, $P(N)$
is the probability, that out of $M=2et_0V/h$ independent attempts $N$
charges are transfered.  Furthermore, Levitov and coworkers studied
the counting statistics of diffusive conductors
\cite{levitov:96-diffusive}, time-dependent problems
\cite{levitov:96-coherent} and of a tunnel junction
\cite{levitov:95-tunnel}.  A theory of full counting statistics based
on the powerful Keldysh-Green's function method was initiated by
Nazarov \cite{yuli:99-annals}. This formulation allows a
straightforward generalization to systems containing superconductors
\cite{belzig:01-super,belzig:01-diff} and multi-terminal structures
\cite{nazarov:02-multi,boerlin:02}.  Classical approaches to FCS were
recently put forward for Coulomb blockade systems
\cite{dejong:96,bagrets:03-coulomb}, and, for chaotic cavities based
on a stochastic path-integral approach \cite{pilgram:03-stochastic}.
The field of counting statistics in the quantum regime is closely
related to the fundamental measuring problem of quantum mechanics,
which has been addressed in a number of works
\cite{levitov:96-coherent,makhlin:00-readout,nazarov:01-measuring,shelankov:03,klich}. 
%% First addressed by Levitov and coworkers \cite{levitov:96-coherent},
%% the problem has been put on firm ground by Nazarov and coworkers
%% \cite{nazarov:01-measuring}. Problems related to the
%% interpretation have been addressed by Shelankov and Rammer
%% \cite{shelankov:03} and by Klich \cite{klich}. 
Expressing the FCS of charge transport by the counting statistics of
photons emitted from the conductor provides an interesting alternative
to classical counting of electrons \cite{beenakker:01-photon}.
Counting statistics has been addressed by now for many different
phenomena
\begin{itemize}
\item Andreev contacts \cite{muzykantskii:94} 
\item generic quantum conductors
  \cite{dejong:96,blanter:01,levitov:01-cumulant,gutman:02-thirdcumulant} 
\item adiabatic quantum pumping 
  \cite{andreev:00-pump,levitov:01-pump,makhlin:01-pump,muzykanstkii:02-pump}
\item qubit-readout
  \cite{makhlin:00-readout,choi:02,engel:02,clerk:03-resonantcooperpair}
\item superconducting contacts in equilibrium \cite{belzig:01-super}
\item proximity effect structures
  \cite{belzig:01-diff,samuelsson:03-counting,reulet:03-diff,belzig:03-incoherent,bezuglyi:03-interferometer}
\item cross-correlations % in multi-tunnel junction structures
  with normal \cite{yuli:01-multi} or
  superconducting contacts \cite{boerlin:02,samuelsson:02-cavity}
\item entangled electron pairs \cite{taddei:02-entanglerfcs,taddei:03-clauserhorne}.
\item phonon counting \cite{kindermann:02-phonon}
\item relation between photon counting and electron counting
  \cite{kindermann:02-photoncounting}
\item current biased conductors
  \cite{kindermann:03-voltage} 
\item interaction effects: weak and strong Coulomb blockade
  \cite{bagrets:03-coulomb,bagrets:03-weakcoulomb,kindermann:03-coulomb}.
\item multiple Andreev reflections in superconducting contacts \cite{cuevas:03-marfcs,johansson:03-marfcs} 

\end{itemize}
Very recently, an important experimental step forward was achieved.
Reulet, Senzier, and Prober measured for the first time the third
cumulant of current fluctuations produced by a tunnel junction
\cite{reulet:03-thirdcumulant}. Surprisingly the measured voltage
dependence deviated from the expected voltage-independent third
cumulant of a simple tunnel contact
\cite{levitov:93-fcs,levitov:01-cumulant}. A subsequent theoretical
explanation is that the third cumulant is in fact susceptible to
environmental effects \cite{beenakker:03-thirdcumulant}. This
experiment has already triggered some theoretical activity
\cite{gutman:02-thirdcumulant,pilgram:03-thirdcumulant,galaktionov:03-thirdcumulant}.

\section{Full Counting Statistics}
\label{sec:fcs}

The fundamental quantity of interest in quantum transport is the
probability distribution
\begin{equation}
  P_{t_0}(N_1,N_2,\ldots,N_M)\equiv P(\vec N)\,,
\end{equation}
which denotes for a $M$-terminal conductor the probability that during a
certain period of time $t_0$ $N_1$ charges enter through terminal 1,
$N_2$ charges enter through terminal 2, $\ldots$, and $N_M$ charges
enter through terminal $M$ (negative $N_i$ correspond to charges
leaving the respective terminal).  The same information is contained
in the cumulant generating function (CGF), defined by
\begin{equation}
  \label{eq:cgf}
  S(\chi)=\ln\left[\sum\nolimits_{\vec N} e^{i\vec N\vec \chi} P(\vec N)\right]\,,
\end{equation}
where we introduced the vector of counting fields $\vec
\chi=(\chi_1,\chi_2,\ldots,\chi_N)$.  The normalization condition
requires $\sum_{\vec N} P(\vec N)=1\,\leftrightarrow\,
  S(\vec \chi=\vec 0)=0$.

\subsubsection{Charge conservation}

We are interested in the long-time limit of the charge counting
statistics, which means that no extra charges remain inside the
conductor after the counting interval. If we count only the total
number of transfered charges, we simply have to consider
$P(N)=\sum_{\vec N} \delta_{\sum N_\alpha,N} P(\vec \chi)$, or,
equivalently, to put all counting fields equal $S(\chi_1=\chi ,
\chi_2=\chi,\ldots,\chi_N=\chi)$. Charge conservation now means that
$S(\chi_1=\chi , \chi_2=\chi,\ldots,\chi_N=\chi)=0$. As a consequence
the CGF depends only on differences between counting fields.  This has
the direct interpretation, that a difference $\chi_\alpha-\chi_\beta$
is related to a charge transfer between terminal $\alpha$ and $\beta$.
In general, this means that we need only $M-1$ counting fields to
describe a $M$-terminal structure.  If one of the counting fields,
e.~g. $\chi_M$, has been eliminated, the charge transfer into terminal
$M$ can be restored from the CGF, in which all other $\chi_\alpha$ are
equal $\chi_\alpha-\chi_M$. In the special case of a two-terminal
device, the CGF depends only on $\chi\equiv\chi_1-\chi_2$. We denote
this below with $S(\chi)$. Later we will see that the CGF's are in
general \textit{periodic} functions of $\chi$, i.~e.
$S(\chi+2\pi)=S(\chi)$.  This ensures that the total charge transfered
is an integer multiple of the electron charge $e$, which makes sense,
since we are talking about electron transport and want to neglect
transient effects.

However, the interesting question, what the charge of an elementary
event is, can be answered by FCS. Suppose the a CGF has the property
$S(\chi+2\pi/n)=S(\chi)$. Direct calculation shows that
\begin{eqnarray}
  P(Q) = \int \frac{d\chi}{2\pi} e^{-iN\chi+S(\chi)} =
  \left\{
    \begin{array}[c]{lll}
      P_n(Q/n) &,& (Q \textrm{ mod } n) = 0 \\
      0 &,& (Q \textrm{ mod } n) \neq 0 
    \end{array}\right.,
\end{eqnarray}
where $P_n(N)$ is the distribution 
$S_n(\chi)=S(\chi/n)$. The probability distribution vanishes for all
$N$ which are not multiples of $n$, thus the elementary charge
transfer is in units of $n e$, where $e$ is the electron charge.
This has interesting consequences in the context of
superconductivity, in which multiple charge transfers can occur
\cite{muzykantskii:94,cuevas:03-marfcs,johansson:03-marfcs}, or for
fractional charge transfer \cite{levitov:01-cumulant}.

\subsubsection{Correlations}

One commonly addressed question is, if two different events (say the
charges transfered into terminals $\alpha$ and $\beta$) are
independent or not. For independent events the probability
distributions are separable and we find that $\langle N_\alpha^k
N_\beta^l \rangle = \langle N_\alpha^k\rangle\langle N_\beta^l
\rangle$. In terms of the CGF this means that the CGF is the sum of
two terms: one which depends only on $\chi_\alpha$ and a second one,
which depends only on $\chi_\beta$. On the contrary, if the CGF can
not be written as such a sum, the charge transfers in terminals
$\alpha$ and $\beta$ are correlated.

\subsubsection{Special distributions (two terminals)}

If the elementary events are uncorrelated, the probability
distribution is \textit{Poissonian}. With the average number of events
is $\bar N$ we have
\begin{equation}
  \label{eq:poisson}
  P_{Poisson}=\frac{\bar N}{N!}e^{-\bar N} 
  \leftrightarrow
  S(\chi)=\bar N\left(e^{i\chi}-1\right)\,.
\end{equation}
In the context of electron transport we encounter this distribution
mostly for \textit{tunnel junctions} with an almost negligible
transmission probability at low temperatures. Here $\bar N=G_TVt_0/e$
is simply related to the voltage bias and the tunnel conductance.

As second example we consider the binomial (or Bernoulli)
distribution. This is obtained if an event occurs with a probability T
and the number of tries is fixed to $N_0$:
\begin{equation}
  \label{eq:binomial}
  P_{binomal} = \binom{N_0}{N} T^N(1-T)^{N_0-N}
  \leftrightarrow
  S(\chi)=N_0 \ln\left[1+T\left(e^{i\chi}-1\right)\right]\,.
\end{equation}
In some sense this is the most fundamental distribution in quantum
transport: it gives the statistics of a voltage biased single channel
quantum conductor if we identify $N_0=eVt_0/h$.

%% Lately it has been shown that, on the contrary, a current biased
%% conductor leads to the so-called Pascal distribution
%% \begin{equation}
%%   \label{eq:pascal}
%%   P_{Pascal} = \binom{N_0}{N-1} T^N(1-T)^{N_0-N-1}\,.
%% \end{equation}
%% It is also called the waiting-time distribution, because it gives the
%% probability that $N_0$ attempts are needed to obtain $N$ successful
%% events. 

\subsubsection{Special distributions (many terminals)}

For uncorrelated processes the CGF takes the simple form
\begin{equation}
  S(\vec \chi)=\sum_{\alpha,\beta} \bar N_{\alpha,\beta}
  \left(e^{i(\chi_\alpha-\chi_\beta)}-1\right) \,.
\end{equation}
The resulting distribution is just the product of Poisson
distributions, taking into account total charge conservation.  An
important example is a multinomial distribution for $N_0$ independent
attempts, which can have different outcomes with probabilities
$T_\alpha $. It has the form
\begin{equation}
  S(\vec \chi)=N_0\ln\left[1+\sum_{\alpha} 
    T_{\alpha}   
    \left(e^{i\chi_\alpha}-1\right)\right] \,.
\end{equation}

%% An interesting property of a general class of  multivariate distribution can be
%% derived in the following way. Suppose the CGF has the form
%% \begin{equation}
%%   S(\vec \chi)= S_q(-i \ln[1+\Lambda(\vec\chi)])
%%   \quad ,\quad \Lambda(\vec\chi)=\sum_{\alpha} 
%%     T_{\alpha} \left(e^{i\chi_\alpha}-1\right)\,.
%% \end{equation}
%% Then the FCS is given by
%% \begin{equation}
%%   P(\vec N)= \sum_Q P_q(Q) \frac{Q!}{(Q-\sum_\alpha N_\alpha)!}
%%   \prod_\alpha \frac{T_\alpha^{N_\alpha}}{N_\alpha !}\,,
%% \end{equation}
%% where $P_q(Q)$ is the distribution corresponding to $S_q(\chi)$. The
%% full counting statistics has therefore the form of a probability that
%% $Q$ charges enter the system, and then are distributed among the
%% various terminal by multinomial distribution.

\section{Theoretical approach to full counting statistics}
\label{sec:theory}

\subsubsection{General theory}

We will follow here the approach to FCS using the Green's function
technique \cite{yuli:99-annals}.  Quantum-mechanically we define 
the cumulant generating function by
\cite{yuli:99-annals,belzig:01-super,belzig:01-diff,levitov:01-cumulant}
\begin{equation}
  \label{eq:cgf-general}
  e^{S(\chi)} =
  \langle {\cal T}_K e^{-i\frac{1}{2e}\int_{C_K} dt \chi(t) I(t)}
  \rangle\,,
\end{equation}
\begin{figure}[t]
  \centering
  \includegraphics[width=8cm]{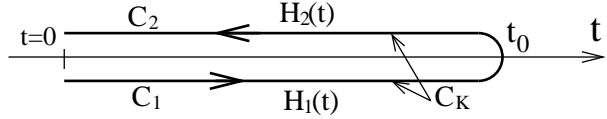}
  \caption{Keldysh time-ordering contour.}
  \label{fig:timepath}
\end{figure}
Here ${\cal T}_K$ denotes time ordering along the Keldysh-contour
$C_K$, depicted in Fig.~\ref{fig:timepath}. The time-dependent field
$\chi(t)$ is defined as $\pm \chi$ for $t\in C_{1(2)}$, i.e. $\chi(t)$
changes sign between the upper and the lower branch of $C_K$. $\hat
I(t)$ is the usual operator of the current through a certain cross
section.  Expansion in the \textit{counting field} yields the
cumulants. In the second order we find the $2^{nd}$ cumulant as
\begin{equation}
  \label{eq:second-cumulant}
  C_2(t_0)=\int_0^{t_0} dt \int_0^{t_0} dt^\prime \left\langle \delta\hat
  I(t)\delta \hat I(t^\prime)\right\rangle\,.
\end{equation}
Higher cumulants yield more complicated expressions.

\subsubsection{Current Correlation Functions}

The cumulants $C_n(t_0)$ are directly related to experimentally
accessible quantities like current noise or the third cumulant of the
current fluctuations. Let us demonstrate the relation for the
low-frequency current noise, defined by
\begin{equation}
  \label{eq:noise}
  S_I=2 \Delta f \int_{-\infty}^{\infty} d\tau
  \left\langle \delta \hat I(\tau) \delta \hat I(0) \right\rangle\,,
\end{equation}
where $\delta \hat I(\tau)=\hat I(\tau)-\langle \hat I \rangle$ and
$\Delta f=f_{max}-f_{min}$ is the frequency band width, in which the
noise is measured. The factor of 2 enters here to conform to the
review article \cite{blanter}.  We now transform in
(\ref{eq:second-cumulant}) the integration variables from $t,t^\prime$
to $T=(t+t^\prime)/2,\tau=t-t^\prime$.  In the limit $t_0\equiv
(\Delta f)^{-1}$ much larger than the correlation time of
current-fluctuations, the integral over $T$ can be evaluated and we
obtain from (\ref{eq:second-cumulant}) the desired result $S_I/2$.
Similar arguments hold for higher cumulants, for which the expression
corresponding to \ref{eq:second-cumulant} are less trivial, however.
In Ref.~\cite{reulet:03-thirdcumulant} it was noted that $C_3$ depends
in an quite unusual way on the frequency band measured, i.e. it is
proportional to $2f_{max}-f_{min}$, which made it possible to prove
experimentally that the third cumulant is actually measured.

\subsubsection{Keldysh-Green's Functions}

So far we have formally defined the CGF quantum mechanically. The
relation to standard quantum-field theory methods is made in the
following way. We introduce the standard Green's function \cite{rammer:86}
in the presence of a time-dependent Hamiltonian 
\begin{equation}
  \label{eq:hpert}
  H_{c}(t) = H_0 + \frac{1}{2e} \chi(t) \hat I \,,
\end{equation}
where the time-dependence is only in the 'counting' field $\chi(t)$. 
The counting field couples to the operator $\hat I$ of the
current through a cross section, which intersects the conductor
entirely. The single-particle operators corresponding to
$H_0$ and $I$ are denoted by $h_0$ and $j$.

Using the matrix notation for the Keldysh-Green's functions, we arrive
at the equation of motion 
\begin{equation}
  \label{eq:eom}
  \left[i\frac{\partial}{\partial t} - \hat h_0 -
    \frac{\chi}{2e} \bar\tau_3 \hat j_c\right] 
  \check G(t,t';\chi) 
  =\delta(t-t^\prime)\,.
\end{equation}
Here $\bar\tau_3$ denotes the third Pauli matrix in the Keldysh space
and is a result of the unusual time-dependence of the counting field.
The relation of the Green's function (\ref{eq:eom}) to the CGF
(\ref{eq:cgf-general}) is obtained from a diagrammatic expansion in
$\chi$ (the calculation is formally equivalent to the calculation of
the thermodynamic potential in an external field, see e.~g. \cite{agd}).  One
obtains the relation \cite{yuli:99-annals}
\begin{equation}
  \label{eq:chi-current}
  \frac{\partial S(\chi)}{\partial \chi} = \frac{it_0}{e} 
  \textrm{Tr}\left[\bar\tau_3\hat j
      \check G(t,t;\chi)\right] \equiv \frac{it_0}{e} I(\chi)\,,
\end{equation}
where we have restricted us to a static situation, for which $\check
G(t,t)$ is independent of time. Note, that the \emph{counting current}
$I(\chi)$ should not be confused with the standard electrical current,
which is actually given by $I_{el}=I(0)$. Rather, $I(\chi)$ contains
(via an expansion in $\chi$) \textit{all current-correlators} at once.
It nevertheless resembles a current in the usual sense. E.~g., it
follows from Eq.~(\ref{eq:eom}) that the counting current is
conserved.

\subsubsection{A simplification}

In a typical mesoscopic transport problem we can access the full
counting statistics based on the separation into terminals (or
reservoirs) and a scattering region. Terminals provide boundary
conditions to Green's function far away from the scattering region.
These are usually determined by external current or voltage sources
and include material properties like superconductivity.  Let
us now take the following parameterization of the current
operator in Eq.~(\ref{eq:eom})
\begin{equation}
  \label{eq:countfield-choice}
  \hat j({\mathbf x})=(\nabla F({\mathbf x}))
  \lim_{{\mathbf x}\to{\mathbf x'}}
  \frac{ie}{2m}\left(\nabla_{\mathbf x}-
    \nabla_{{\mathbf x}'}\right)\hat\sigma_3\,.
\end{equation}
$F({\mathbf x})$ is chosen such that it changes from 0 to 1 across a
cross section C, which intersects the terminal, but is of arbitrary
shape. Here we have introduced a matrix $\hat\sigma_3$ in the current
operator, occurring e.~g. in the context of superconductivity.  We
assume that the change from 0 to 1 should occur on a length scale
$\Lambda$, for which we assume $\lambda_F\ll\Lambda\ll l_{imp},\xi_0$
(Fermi wave length $\lambda_F$, impurity mean free path $l_{imp}$, and
coherence length $\xi_0=v_F/2\Delta$). With this assumption we can
reduce Eq.~(\ref{eq:eom}) \textit{inside the terminal} to its
quasiclassical version (see Ref.~\cite{rammer:86})
\begin{equation}
  \label{eq:eilenberger-counting}
  {\mathbf v_F }\nabla \check g({\mathbf{x}},{\mathbf v_F},t,t',\chi)
  = \left[-i\frac{\chi}{2} (\nabla F({\mathbf x})) {\mathbf v_F} \check \tau_K 
  \,,\,\check g({\mathbf{x}},{\mathbf v_F},t,t',\chi) \right] \,.
\end{equation}
Here $\check\tau_K=\bar\tau_3\hat\sigma_3$ is the matrix of the
current operator and $\check g$ obeys the normalization condition
$\check g^2=1$. Other terms can be neglected due to the assumptions we
have made for $\Lambda$.  The counting field can then be eliminated by
the gauge-like transformation
\begin{equation}
  \check g({\mathbf x},{\mathbf v_F},t,t',\chi) 
  = e^{-i\chi F({\mathbf x})\check \tau_K/2}
  \check g({\mathbf x},{\mathbf v_F},t,t',0) 
  e^{i\chi F({\mathbf x})\check \tau_K/2}\,.
\end{equation}
We assume now that the terminal is a diffusive metal of negligible
resistance. Then the Green's functions are constant in space (except in
the vicinity of the cross section C) and isotropic in momentum space.
Applying the diffusive approximation \cite{rammer:86} in the terminal
leads to a transformed terminal Green's function
\begin{equation}
  \label{eq:countrot}
  \check G(\chi) = e^{-i\chi\check \tau_K/2}
  \check G(0) e^{i\chi\check \tau_K/2}\,,
\end{equation}
on the right of the cross section $C$ (where $F({\mathbf x})=1$) with
respect to the case without counting field.  Consequently, the counting
field is entirely incorporated into a \textit{modified boundary
  condition} imposed by the terminal onto the mesoscopic system.

\subsubsection{Summary of Theoretical Approach}

This concludes the theoretical approach to counting statistics of
mesoscopic transport. Let us briefly summarize the scheme to follow.
The FCS can be obtained by a slight extension of the usual
Keldysh-Green's function approach, which is widely employed to treat
quantum transport problems. Making use of the separation of the
mesoscopic structure into \textit{terminals} and a \textit{scattering}
region, the formalism boils down to a very powerful, but nevertheless
simple rule: we have to apply the \textit{counting rotation}
(\ref{eq:countrot}) to a terminal, thus providing new boundary
conditions (now depending on the \textit{counting field} $\chi$) to
the scattering problem. We then proceed 'as usual' and calculated the
current in the terminal, which again depends on $\chi$. Finally the
counting statistics is obtained from Eq.~(\ref{eq:chi-current}).

\section{Two-Terminal contacts}
\label{sec:qpc}

\subsubsection{Tunnel contact}

To illustrate the theoretical method we first calculate the counting
statistics of a tunnel junction. As usual the system is described by a
tunnel Hamiltonian $H=H_1+H_2+H_T$, where $H_{1(2)}$ describe the
left(right) terminal and $H_T$ describes the tunneling. The current is
calculated in second order in the tunneling amplitudes and we obtain
%% \begin{equation}
%%   \label{eq:tunnel-current2}
%%   I(\chi)=\frac{G_T}{8e}\int dE \textrm{Tr}\left(
%%     \check\tau_K\left[\check G_1(\chi),\check G_2\right]\right)\,,
%% \end{equation}
$I(\chi)=\frac{G_T}{8e}\int dE \textrm{Tr}\left(
    \check\tau_K\left[\check G_1(\chi),\check G_2\right]\right)$,
where $G_T$ is the conductance of the tunnel junction and we have
included the counting field in $\check G_1$. 
The CGF is (using $ (\partial/\partial\chi) %\frac{\partial}{\partial\chi}
G_1(\chi)=(i/2)\left[\check\tau_K,\check G_1(\chi)\right]$)
\begin{equation}
  S(\chi)=i\frac{t_0}{e}\int_0^\chi d\chi^\prime I(\chi^\prime)
  = \frac{G_Tt_0}{4e^2} \int dE \textrm{Tr}\left\{
    \check G_1(\chi),\check G_2\right\}\,,
  \label{eq:cgf-tunnel-general}
\end{equation}
which is the general expression for the FCS of a tunnel junction.  We
use the pseudo-unitarity $\check\tau_K^2=\check 1$ to write
\begin{equation}
  \label{eq:cgf-tunnel}
  S(\chi) = N_{12}(e^{i\chi}-1)+N_{21}(e^{-i\chi}-1)\,,
\end{equation}
where $ N_{ij} = (t_0G_T/16e^2)\int dE \textrm{Tr}
\left[(1+\check\tau_K) \check G_i (1-\check\tau_K)\check G_j\right]$ denotes the
average number of charges tunnel from $i$ to $j$. The statistics is
therefore a bidirectional Poisson distribution \cite{levitov:01-cumulant}.
It is easy to see that the cumulants are
$C_{n}=N_{12}+(-1)^n N_{21}$. % and $C_{2n}=N_{12}+N_{21}$.
If either $N_{21}=0$ or $N_{12}=0$ we obtain the Schottky
limit. Furthermore, in equilibrium $N_{12}=N_{21}$ and the FCS is
$(2G_Tk_BT t_0/e^2)(\cos(\chi)-1)$, which is non-Gaussian, remarkably.

\subsubsection{General CGF for quantum contacts}

Using the method presented in the previous section, we can find the
counting statistics for all conductors, which are characterized by a
set of transmission coefficients $\{T_n\}$. Nazarov has shown that the
transport properties of such a contact are described by a {\em matrix
  current} \cite{nazarov:99-circuit}
\begin{equation}
  \label{eq:matrix-current}
  \check{I}_{12}=-\frac{e^2}{\pi}\sum_n 
  \frac{2T_n\left[\check G_1,\check G_2\right]}{
    4+T_n\left(\{\check G_1,\check G_2\}-2\right)}\,.
\end{equation}
Here $\check G_{1(2)}$ denote the matrix Green's functions on the left
and the right of the contact. We should emphasize that the matrix form
of (\ref{eq:matrix-current}) is crucial to obtain the FCS, since it is
valid for any matrix structure of the Green's functions. The
\textit{scalar current} is obtained from the matrix current by
\begin{equation}
  \label{eq:el-current}
  I_{12}=\frac{1}{4e}\int dE {\rm Tr} \check\tau_K\check I_{12}\,.
\end{equation}
To find the FCS, we apply the \textrm{counting rotation}
(\ref{eq:countrot}) to terminal 1, i.~e. $\check G_1$ becomes
$\chi$-dependent. It turns out that the CGF can then be found
generally from the relations (\ref{eq:chi-current}),
(\ref{eq:matrix-current}), and (\ref{eq:el-current}). To integrate
(\ref{eq:chi-current}) with respect to $\chi$, we need the relations
$i(\partial/\partial\chi)\check G_1(\chi)=[\check\tau_K,\check
G_1(\chi)]$ and $\textrm{tr}[\check G_1(\chi),\{\check
G_1(\chi),\check G_2\}^n]=0$. We find \cite{belzig:01-super}
\begin{equation}
  \label{eq:cgf-two-terminal}
  S(\chi)=\frac{t_0}{2\pi}\sum_n\int dE {\textrm{Tr}}
  \ln\left[1+\frac{T_n}{4}
    \left(\{\check G_1(\chi),\check G_2\}-2\right)\right]\;.
\end{equation}
This is a very important result. It shows that the counting statistics
of a large class of constrictions can be cast in a common form,
independent of the contact types. 
Note, that Eq.~(\ref{eq:cgf-two-terminal}) is just the sum over CGF's
of all eigenchannels. Thus, we can obtain the CGF's of all
constrictions from a known transmission eigenvalue density. These are
known for a number of generic contacts (see e.g.
\cite{beenakker:97-rmp} and Table~\ref{tab:cgf}), can be determined
numerically, or can be taken from experiment.  Below we will 
discuss several illustrative examples for a single channel contacts.

\begin{table}[tbp]
  \centering
  \begin{tabular}[t]{r||c|c||}
    & $\displaystyle\rho(T) [G/G_Q] $ & 
    $\displaystyle \check s(\Lambda )$ \\\hline\hline
    Single channel & $\displaystyle\delta(T-T_1)$ & $\displaystyle\ln(1-T_1(\Lambda-1)/2)$ \\\hline
    Diffusive connector  & $ \displaystyle\frac{1}{2}\frac{1}{T\sqrt{1-T}}$ & 
    $\displaystyle\frac{1}{4}\textrm{acosh}^2(\Lambda)$ \\\hline
    Dirty interface & $\displaystyle\frac{1}{\pi}\frac{1}{T^{3/2}\sqrt{1-T}}$
    & $\displaystyle \sqrt{2(1+\Lambda)}$ \\\hline
    Chaotic cavity & $\displaystyle \frac{2}{\pi}\frac{1}{\sqrt{T}\sqrt{1-T}} $
    & $\displaystyle 4 \ln\left(2+\sqrt{2(1+\Lambda)}\right)$ \\\hline\hline
  \end{tabular}
  \caption{Characteristic functions of some generic conductors. The
    transmission eigenvalue densities are normalized to $G/G_Q$, where
    $G_Q=2e^2/h$ is the quantum conductance. The third column displays
    the CGF-density, which determines the CGF via $S(\chi)=(t_0G/4eh) \int
    dE \textrm{tr}\check s(\{\check G_1(\chi),\check G_2\}/2)$.}
  \label{tab:cgf}
\end{table}

\subsubsection{Normal contacts}

Consider first a single channel with transmission T between two normal
reservoirs. They are characterized by occupation factors $f_{1(2)} =
[\exp((E-\mu_{1(2)})/k_B T_e)+1]^{-1}$ ($T_e$ is the temperature). We
obtain the result \cite{levitov:93-fcs,levitov:96-coherent} (see
Appendix)
\begin{eqnarray}
  \label{eq:cgf-normalcontact}
  S(\chi)=\frac{2t_0}{h}\int dE
  \ln\left[1+T_{12}(E)\left(e^{i\chi}-1\right)
    +T_{21}(E)\left(e^{-i\chi}-1\right)\right]\,.
\end{eqnarray}
Here we introduced the probabilities $T_{12}=T
f_1(E)\left(1-f_2(E)\right)$ for a tunneling event from 1 to 2 and
$T_{21}(E)$ for the reverse process. We see that the FCS (for each
energy) is a trinomial of an electron going from left to right, from
right to left, or no scattering at all.  The {\em counting factors}
$e^{\pm i\chi}-1$ thus correspond to single charge transfers from 1 to
2 (2 to 1).

At zero temperature and $\mu_1-\mu_2=eV\ge 0$ the argument of the
energy integral is constant in the interval $\mu_1<E<\mu_2$ and we
obtain the \textit{binomial form} $S(\chi)=\frac{2et_0|V|}{h}
\ln\left[1+T \left(e^{i\chi}-1\right)\right]$. Note that for reverse
bias $\mu_2>\mu_1$ the CGF has the same form, but with a counting
factor $e^{-i\chi}-1$.  The prefactor denotes the \emph{number of
  attempts} $M=e t_0V/h $ to transfer an electron \footnote{The
  noninteger values of $M(t_0)$ occur due to the quasiclassical
  approximation \cite{levitov:96-coherent}. A more careful treatment
  reveals that $M$ itself is described by a probability distribution.
  For large $M$ the difference is negligible.}.  If the transmission
probability is unity the FCS is non-zero only for $N=M$, which
therefore constitutes the maximal number of electrons occupying an
energy strip $eV$ that can be sent through one (spin-degenerate)
channel in a time interval $t_0$. In equilibrium it follows from
Eq.~(\ref{eq:cgf-normalcontact}) that the counting statistics is
\cite{levitov:03-noisebook}
\begin{equation}
  \label{eq:cgf-equilibrium}
  S(\chi) = -\frac{2 t_0 k_B T_{el}}{h} 
  \textrm{asin}^2\left(\sqrt T \sin\frac{\chi}{2}\right)\,.
\end{equation}
The fluctuations are non-Gaussian, except for $T=1$, when
$S(\chi)=-\frac{t_0 k_B T_{el}}{h}\chi^2$.  

\subsubsection{SN-contact}

The FCS of a contact between a superconductor and a normal metal also
follows from the general expression Eq.~(\ref{eq:cgf-two-terminal}).
Using the Green's functions given in the Appendix we find the result
\cite{muzykantskii:94}
\begin{equation}
  \label{eq:cgf-sncontact}
   S(\chi)=\frac{t_0}{2\pi}\sum_n\int dE
  \ln\left[1+\sum\limits_{q=-2}^{2} 
    A_{nq}(E)\left(e^{iq\chi}-1\right)\right]\,.
\end{equation}
The coefficients $A_{nq}(E)$ are related to a charge transfer of
$q\times e$.  For example, a term $\exp(2i\chi)-1$ corresponds just to
an Andreev reflection process, in which two charges are transfered
simultaneously.  Explicit expressions for the various coefficients are
given in Refs. \cite{muzykantskii:94,belzig:03-book}. The most
interesting regime is that of pure Andreev reflection:
$eV,k_BT\ll\Delta$. Here we obtain
\begin{equation}
  \label{eq:cgf-andreev-general}
  S(\chi)=\frac{t_0}{h}\int dE 
  \ln\left[1+R_A f_+f_-\left(e^{i2\chi}-1\right)+
    R_A(1-f_+)(1-f_-))\left(e^{-i2\chi}-1\right)\right]\,,
\end{equation}
where $R_A=T^2/(2-T)^2$ is just the Andreev reflection probability and
$f_\pm=f(\pm E)$ denotes the occupation with electrons above(below)
the chemical potential of the superconductor.  For low temperatures
$k_BT_{e}\ll eV \ll\Delta$, the CGF becomes
\begin{equation}
  S(\chi)=\frac{2et_0 |V|}{h} 
  \ln\left[1+R_A\left(e^{i2\chi}-1\right)\right]\,.
  \label{eq:cgf-andreev}
\end{equation}
The CGF is now $\pi$-periodic, which according to Sec.~\ref{sec:fcs}
reflects that the charge transfer of an elementary event is now $2e$,
a consequence of Andreev reflection. Quite remarkably, the statistics
is again a simple binomial distribution.
In equilibrium, we can adapt the result from
Eq.~\ref{eq:cgf-equilibrium} to find 
\begin{equation}
  S(\chi) = -\frac{2 t_0 k_B
    T_{el}}{h} \textrm{asin}^2\left(\sqrt{R_A}\sin\chi\right)\quad
  (\textrm{for} \chi\in[-\pi/2,\pi/2])\,.
\end{equation}
The counting statistics is also non-Gaussian, except for $R_A=1$.

\subsubsection{Superconducting Contact}
\label{sec:mar}

Now we turn to a slightly more involved problem: a contact between two
superconductors biased at a finite voltage $V$. For $eV < 2 \Delta$
the transport is dominated by multiple Andreev reflections (MAR).  The
microscopic analysis of the average current and the shot noise
calculations suggest that the current at subgap energies proceeds in
``giant" shots, with an effective charge $q \sim e(1 + 2\Delta/|eV|)$.
However, the question of size of the charge transfered in an
elementary event can only be rigorously resolved by the FCS. The
answer was given by Cuevas and the author \cite{cuevas:03-marfcs}
based on a microscopic Green's function approach.  Independently,
Johansson, Samuelsson and Ingerman \cite{johansson:03-marfcs} arrived
at the same conclusion using a different method.

Now, what would we like to have? In Sec.~\ref{sec:fcs} we have
discussed that one can speak of multiple charge transfers if the CGF
allows an interpretation in terms of elementary events, which are
described by counting factors $e^{in\chi}-1$, where $n$ denotes the
charge transfered in the process. How can we ever hope to obtain this
from the general formula~(\ref{eq:cgf-two-terminal})? We have to
calculate the determinant of a 4$\times$4-matrix, which can give only
factors of the type $e^{i2\chi}$ or even smaller charges. The answer
to this puzzle is that we have to reinterpret the matrix structure in
(\ref{eq:cgf-two-terminal}), since the Green's functions of
superconductors at a finite bias voltage are essentially nonlocal in
energy. The general result for the CGF can be written as
$S(\chi)=\textrm{Tr}\ln \check Q$, where Tr$=\int_0^{t_0} dt$tr and
$\check Q(t)=1+(T/4)(\{\check G_1\stackrel{\otimes}{,}\check
G_2\}-2)(t,t)$. Here $\check G_1\otimes\check
G_2(t,t^\prime)=\int dt^{\prime\prime} \check G_1(t,t^{\prime\prime}\check
G_2(t^{\prime\prime},t^\prime)$. Let us set the chemical
potential of the right electrode to zero and represent the Green's
functions by $\check G_1(t,t^{\prime}) = e^{i eVt \bar \tau_3} \check
G_S(t-t^{\prime}) e^{-i eVt^\prime \bar \tau_3}$ and $\check
G_2(t,t^{\prime}) = \check G_S(t-t^{\prime})$. Here, we have not
included the dc part of the phase, since it can be shown that it drops
from the expression of the dc FCS at finite bias.  The Fourier
transform leads to a representation of the form $\check
G(E,E^{\prime}) = \sum_{n} \check G_{0,n}(E) \delta(E - E^{\prime}
+neV)$, where $n=0,\pm 2$.  Restricting the fundamental energy
interval to $E-E^\prime \in [0,eV]$ we can represent the convolution
as \textit{matrix product}, i.e.  $(G_1 \otimes G_2) (E,E^{\prime})
\to (\check G_1 \check G_2)_{n,m} (E,E^{\prime}) = \sum_k
(G_1)_{n,k}(E,E^\prime) (G_2)_{k,m}(E,E^{\prime})$.
The trace in this new representation is written as $\int_0^{eV} dE
\sum_n \textrm{Tr} \ln \left(\check Q\right)_{nn}$. In this way, the
functional convolution is reduced to matrix algebra for the
infinite-dimensional matrix $\check Q$.  From these arguments it is
clear that the statistics is a \textit{multinomial} distribution of
\textit{multiple} charge transfers:
\begin{equation}
  \label{eq:marfcs}
  S(\chi) = \frac{t_0}{h}\int_0^{eV} dE \ln \left[
  1+\sum_{n=-\infty}^{\infty}P_n(E,V)\left(e^{in\chi}-1\right)\right]\,.
\end{equation}
General expressions for the probabilities $P(E,V)$ have been derived
in Ref.~\cite{cuevas:03-marfcs}. 

Here we will pursue a different path and study a toy model.  Let us
neglect all set $f^{R,A}(|E|<\Delta)=1$, $g^{R(A)}(|E|>\Delta)=\pm 1$,
and equal to zero otherwise. Physically, this means that we neglect
Andreev reflections above the gap and replace the quasiparticle
density of states by a constant $|E|>\Delta$. This simplifies the
calculation a lot, since the matrix trace now becomes finite. Let us
for example consider a voltage $eV=2\Delta/4$. In that case, we
consider the determinant of the matrix
\begin{equation}
  \textrm{det}\left[1-\frac{\sqrt{T}}{2}
  \left(
    \begin{array}[c]{ccccc}
      \hat Q_-(\chi) & 1 \\
      1 & 0 & e^{- i\chi\hat\tau_3} \\
      &  e^{i\chi\hat\tau_3} & 0 & 1\\
      & & 1 & 0 & e^{- i\chi\hat\tau_3} \\
      & & & e^{i\chi\hat\tau_3} & \hat Q_+\\
    \end{array}\right)\right]\,,
\end{equation}
where $ Q_\pm(\chi)$ describe quasiparticle emission (injection) and
off-diagonal pairs $e^{\pm\chi}$ are associated with Andreev
reflection. Evaluating the determinant we find $S(\chi)=\frac{\Delta
  t_0}{2h} \ln\left[1+P_5\left(e^{in\chi}-1\right)\right]$, where
$P_5=T^5/(16-20T+5T^2)^2$. This expression describes binomial
transfers of $5$ charges with probability $P_5$. For general
subharmonic voltages $2\Delta/(n-1)$ we find
\begin{equation}
   S(\chi)=\frac{2\Delta t_0}{(n-1)h} 
      \ln\left[1+P_n\left(e^{in\chi}-1\right)\right]\,,
\end{equation}
where the probabilities are given by
\begin{equation}
  \begin{array}{l}
    P_2 = \frac{T^2}{(2-T)^2} \,,\,
    P_3 = \frac{T^3}{(4-3T)^2} \,,\,
    P_4 = \frac{T^4}{(8-8T+T^2)^2}\,,\,
    P_5 = \frac{T^5}{(16-20T+5T^2)^2}\\
    P_6 = \frac{T^6}{(2-T)^2(16-16T+T^2)^2}\,,
    P_7 = \frac{T^7}{(64-112T+56T^2-7T^3)^2}\,.
%    P_8 = {\frac {{T}^{8}}{ \left( {T}^{4}-32\,{T}^{3}+160\,{T}^{2}-256\,T+128
%        \right) ^{2}}}
  \end{array}
\end{equation}
Note the limiting cases of these probabilities $P_n\sim T^n/4^{n-1}$
for $T\ll 1$ and $P_n=1$ for $T\to 1$. We conclude this section by
saying that the general results for the CGF \cite{cuevas:03-marfcs}
allow for a fast and efficient calculation of all dc-transport
properties of contacts between superconductors (which may contain
magnetic impurities, phonon broadening or other imperfections).

\section{Quantum Noise in Diffusive SN-Structures}
\label{sec:diff}

\begin{figure}[tb]
  \centering
  \includegraphics[width=0.9\textwidth,keepaspectratio,clip]{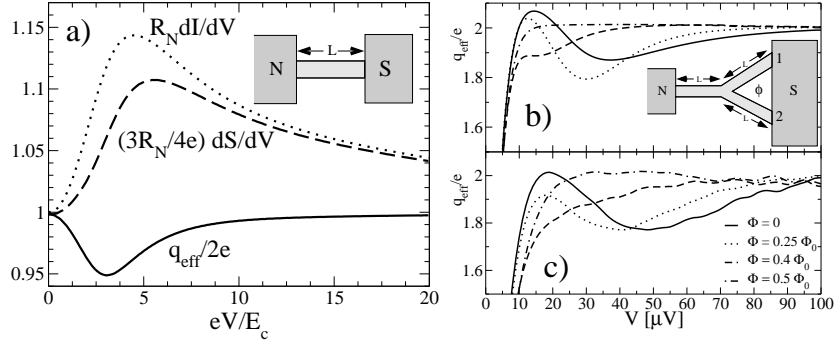}
  \caption{Noise in diffusive SN-systems. Left panel a):
    the differential conductance and the noise show a reentrant
    behavior. The effective charge, defined as
    $q_{eff}(E)=(3/2)dS/dI$ reveals that the correlated Andreev pair
    transport suppresses the noise below the uncorrelated
    Boltzmann-Langevin result $2e$.  Right panels b) and c): Effective
    charge of the Andreev interferometer shown in the inset (realized
    experimentally in Ref.~\cite{reulet:03-diff}).  The upper panel b)
    shows the theoretical predictions and the lower panel c) the
    experimental results. The theoretical results contain no fitting
    parameter (the Thouless energy $E_c=30\mu$eV was extracted from
    the sample geometry and the experimental temperature of $43$mK was
    included in the calculation). Therefore, it is reasonable that
    the deviations between experimental and theoretical results comes
    from possible heating effects in the experiment, which are not
    accounted for in the theoretical calculation.}
  \label{fig:diff}
\end{figure}

In this section, we illustrate a further advantage of the
Keldysh-Green's functions approach to counting statistics. We consider
a normal metallic diffusive wire connected on one end to a normal
metal reservoir and on the other side to a superconductor. The wire is
supposed to have a mean free path $l\gg\lambda_F$, a corresponding
diffusion coefficient $D=v_F l/3$, and a length $L$. For
$eV,k_BT\ll\Delta$ the transport occurs through Andreev reflection at
the interface to the superconductor. This system shows a quite
remarkable property, which is the so-called reentrance effect of the
conductance.  The energy difference $2E$ of electron-hole pairs leads
to a dephasing on a length scale $\xi_E=\sqrt{D/2E}$. This has the
consequence that the (otherwise) normal wire becomes partially
superconducting and the conductance increases with decreasing energy.
However, once the coherence length $\xi_E$ reaches $L$ the conductance
\textit{decreases} again. Finally for $E=0$ the conductance is
\textit{exactly} equal to the conductance in the normal state. This is
the reentrance effect occurring at an energy of the order of the
Thouless energy $E_c=\hbar D/L^2$. In Fig.~\ref{fig:diff} (left panel,
dotted curve) the resulting differential conductance at zero
temperature is plotted.

The transport in this system is described by a matrix diffusion
equation for the Keldysh Green's functions, the so-called Usadel
equation 
\begin{equation}
  \label{eq:usadel}
  -\frac{D}{\sigma}\nabla \check I =
  \left[ -i E\hat\tau_3 , \check G \right]\;,\;
  \check I = -\sigma  \check G \nabla\check G\,.
\end{equation}
In these equations $\sigma=2e^2N_0D$ is the conductivity. The boundary
conditions for this equation are that the Green's functions in the
terminal approach the bulk solution $\check G_N$ or $\check G_S$,
respectively. This equation is in general difficult to solve, even if
one is interested in the average current only. However, we can
calculate the noise and the counting statistics using the recipe
outlined in Sec.~\ref{sec:theory} and obtain the noise in the full
parameter range of Eq.~(\ref{eq:usadel}).

Before considering Eq.~(\ref{eq:usadel}) in its full generality, we
consider the limiting cases of low and high energies (compared to
$E_c$). For $E=0$ the r.h.s. is absent and the system is completely
analogous to a diffusive connector as discussed in \ref{sec:qpc}. From
Table~\ref{tab:cgf} and using the eigenvalues (\ref{eq:sneigenvalues})
we find
%% \begin{eqnarray}
%%   \label{eq:cgf-diffusive-andreev}
%%   S(\chi) & = & \frac{t_0 G}{16e^2} \int dE \times  \\\nonumber
%%   &&\textrm{acosh}^2\left[2\left(f_+f_-(e^{2i\chi}-1) + 
%%     (1-f_+)(1-f_-)(e^{-2i\chi}-1)\right)-1\right]\,.
%% \end{eqnarray}
\begin{equation}
  \label{eq:cgf-diffusive-andreev}
  S(\chi) = \frac{t_0 G}{16e^2} \int dE\textrm{acosh}^2\left[2\left(f_+f_-(e^{2i\chi}-1) + 
    (1-f_+)(1-f_-)(e^{-2i\chi}-1)\right)-1\right].
\end{equation}
This result shows, once again, that the charges are transfered in
pairs. It is interesting to compare with the CGF for a diffusive wire
between two normal metals, for which we obtain
\cite{levitov:96-diffusive,yuli:99-annals}
\begin{equation}
  \label{eq:cgf-diffusive-normal}
  S(\chi) = \frac{t_0 G}{4e^2} \int dE
  \textrm{acosh}^2\left[2\left(f_1(1-f_2)(e^{i\chi}-1) + 
    f_2(1-f_1)(e^{-i\chi}-1)\right)-1\right].
\end{equation}
We see that the only difference in the CGF between the SN- and the
NN-case is in the counting factors, and a prefactor $1/4$. Note, that
this coincidence only occurs for the diffusive connector, but is by no
means a general rule. At zero temperature the results simplify and we
find
\begin{equation}
  \label{eq:cgf-diff-zerotemp}
  S^{\textrm{SN}}(\chi)=\frac12 S^{\textrm{NN}}(2\chi) \; , 
  \;
  S^{\textrm{NN}}(\chi)=\frac{t_0 G V}{4e} \textrm{acosh}^2\left(2e^{i\chi}-1\right)\,,
\end{equation}
a surprising simple relation between the CGF for the Andreev wire and
the normal diffusive wire. It is easy to see that the cumulants obey
the general relation $ C_n^{\rm SN}=2^{n-1} C_n^{\rm NN}$. We observe
that we can read off the effective charge from the ratio $C_n^{\rm
  SN}/C_n^{\rm NN}$ $=$ $(q_{eff}/e)^{n-1}$ and, indeed, find
$q_{eff}=2e$.  This result for the effective charge is a special
property of the \emph{diffusive connector}.

At energies large compared to $E_c$ it is also possible to find the
CGF for the Andreev wire in general. Then the proximity effect in the
wire is absent and it turns out \cite{belzig:03-incoherent} that the
wire can be effectively mapped on a normal circuit, consisting of two
identical wires in series to which twice the voltage is applied and
twice the counting field. Thus, for $E \gg E_c$ we obtain
$S^{SN}(\chi)$ from $S^{NN}(\chi)$ by the replacement $\chi\to2\chi$
and $G\to G/2$, which exactly brings us to
Eq.~\ref{eq:cgf-diffusive-andreev} and shows that the counting
statistics is again the same in the incoherent limit.

The full quantum-mechanical calculation of the energy-dependent shot
noise can be performed on the basis of the approach of
Sec.~\ref{sec:theory} \cite{belzig:01-diff}.  We expand up to linear
order in $\chi$, i.e. $\check G(\chi)=\check G_0-i(\chi/2) \check
G_1$ and $\check I(\chi)=\check I_0-i(\chi/2) \check I_1$.
Substituting in (\ref{eq:usadel}) we find
\begin{equation}
  \frac{D}{\sigma}\frac{\partial}{\partial x}\check I_1  = 
  \left[-iE\bar\tau_3\,,\,\check G_1\right]\,,\, 
  \check I_1 = 
  - \sigma \left( \check G_0\frac{\partial}{\partial x}\check G_1
    + \check G_1\frac{\partial}{\partial x}\check G_0\right).
  \label{eq:usadelnoise}
\end{equation}
The boundary conditions at the reservoirs read $\check
G_1(0)=\left[\check\tau_{\text{K}},\check G_{\text{L}}\right]$ at the
left end and $\check G_1(L)=0$ at the right end.  Finally the noise is
$S_{\text{I}}=-e\int dE \text{Tr}\check\tau_{\text{K}}\check I_1(x)$.
By taking the trace of Eq.~(\ref{eq:usadelnoise}) multiplied with
$\check\tau_{\text{K}}$ it follows that it does not matter, where the
noise is evaluated, as it should be. From these equations the
generalization of the Boltzmann-Langevin equation to superconductors
can be derived \cite{pistolesi:03}, which allows for a faster
numerical solution. The results for the energy dependent noise is
shown in the left panel Fig.~\ref{fig:diff}. A direct comparison of
the differential shot noise and the differential conductance (for zero
temperature) shows the difference in the energy dependence. The
effective charge defined as $q_{eff}=(3/2) dS/dI$ displays the clear
deviation of the quantum noise from the Boltzmann-Langevin result of
$2e$. At energies below the Thouless energy $E_c$ the
effective charge is suppressed below $2e$. This shows that the
correlated Andreev pair transport suppresses the noise below the
uncorrelated Boltzmann-Langevin result.

To experimentally probe the pair correlations in diffusive
super\-con\-duc\-tor-normal metal-hetero\-structures it is most
convenient to use an Andreev interferometer. An example is shown in
the left part of Fig.~\ref{fig:diff}. A diffusive wire connected to a
normal terminal is split into two parts, which are connected to two
different points of a superconducting terminal. By passing a magnetic
flux through the loop one can effectively vary the phase difference
between the two connections to the superconductor. Such a structure
has been experimentally realized by the Yale group \cite{reulet:03-diff}.
In Fig.~\ref{fig:diff} we present a direct comparison between our
theoretical predictions and the experimentally obtained effective
charge. Note, that we have included the experimental temperature in the
theoretical modeling.  The finite temperature explains the strong
decrease of the effective charge in the regime $|eV|\leq k_BT$, where
the noise is fixed by the fluctuation-dissipation theorem. The
disagreement between theory and experiment in this regime stems solely
from differences in the measured temperature-dependent conductance
from the theoretical prediction. We attribute this to heating effects.
The qualitative agreement in the shot-noise regime $|eV|\geq k_BT$ is
satisfactory, if one takes into account, that we have no free
parameters for the theoretical calculation. Both, experiment and
theory show a suppression of the effective charge for some finite
energy, which is of the order of the Thouless energy and depends on
flux in a qualitative similar manner. Remarkably for half-integer flux
the effective charge is completely flat, in contrast to what one would
expect from circuit arguments based on the conductance distribution in
the fork geometry. Currently we have no explanation for this
behavior, and therefore more work is needed in this direction.

\section{Multi Terminal Circuits}
\label{sec:split}

In circuits with more than two terminals it is of particular interest
to study non-local correlations of currents in different terminals.
For that purpose we need a slight extension of the theoretical
approach of Sec.~\ref{sec:theory}, suitable for multi-terminal
circuits. We will now introduce this method.

\subsubsection{Circuit Theory}
\label{sec:circuit}

To study transport in general mesoscopic multi-terminal structures the
so-called circuit theory for quantum transport was developed by
Nazarov \cite{nazarov:99-circuit,yuli:94-circuit}.  Its main idea,
borrowed from Kirchhoff's classical circuit theory, is to represent a
mesoscopic device by discrete elements, which resemble the known
elements of electrical transport. We briefly repeat the essentials of
the circuit theory. Topologically, one distinguishes three elements:
terminals, nodes and connectors.  Terminals are the connections to the
external voltage or current sources and provide boundary conditions,
specifying externally applied voltages, currents or phase differences
in the case of superconductors.  The actual circuit is represented by
a network of nodes and connectors, the first determining the
approximate layout and the second describing the connections between
different nodes, respectively.

To describe quantum effects it is necessary to represent the variables
describing a node by \textit{matrix Green's function} $\check G$,
which can be either Nambu or Keldysh matrices, or a combination
thereof. Consequently, we describe the current through a connector by
a \textit{matrix current} $\check I$, which relates the fluxes of all
elements of $\check G$ on neighboring nodes. The current has been
derived by Nazarov \cite{nazarov:99-circuit} and is given by
Eq.~(\ref{eq:matrix-current}) for a connector, characterized by a set
of transmission coefficients $\{T_n\}$. Note that the \emph{electrical
  current} is obtained from $ I_{12}=\frac{1}{4e}\int dE {\rm Tr}
\check\tau_K\check I_{12}$. The boundary condition are given in terms
of fixed matrix Green's functions $\check G_i$, which are determined
by the applied potential, the temperature, the type of lead, and a
counting field $\chi_i$.

Once the network is determined and all connectors are specified, the
transport properties can be found by means of the following circuit
rules.  We associate an (unknown) Green's function $\check G_j$ to
each node $j$. The two rules are
\begin{enumerate}
\item $\check G^2_j=\check 1$ for the Green's functions of all internal
  nodes $j$.
\item the total matrix current in a node is conserved: $\sum_i \check
  I_{ij}=0$, where the sum goes over all nodes or terminals connected to
  node $j$ and each matrix current is given by
  (\ref{eq:matrix-current}). 
\end{enumerate}
Finally, the observable currents into the terminals are given by
$I_i=\sum_j I_{ij}$, where the sum runs over all nodes connected to
the terminal $i$. To obtain the counting statistics, we finally
integrate all currents
$I_i(\vec\chi)=(\partial/\partial\chi_i)S(\vec\chi)$ to find the CGF
$S(\vec\chi)$.

\begin{figure}[tb]
  \centering
  \includegraphics[width=8cm,keepaspectratio,clip]{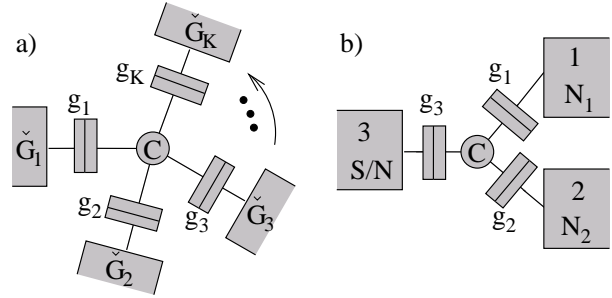}
  \caption{Multi tunnel junction structure: a) general setup with K
  terminals connected to a common node. b) beam splitter setup in
  which terminal 3 is either a normal metal or a superconductor.}
  \label{fig:multitunnel}
\end{figure}

\subsubsection{Multi tunnel junction structure}

A general expression of $S(\vec\chi)$ can be obtained for a system of
an arbitrary number of terminals connected to one common node by
tunnel contacts, see Fig.~\ref{fig:multitunnel}
\cite{yuli:01-multi,boerlin:02}. At the same time it nicely
demonstrates the application of the circuit theory rules, presented
above.  Let us denote the unknown Green's function of the central node
by $\check G_c(\vec\chi)$. The matrix current from a terminal $\alpha$
($\alpha=1,\ldots,K$) into the central node is given by the relation
\begin{equation}
 \check I_\alpha(\vec\chi) = \frac{g_\alpha}{2} \left[ \check
 G_c(\vec\chi) , \check G_\alpha(\chi_\alpha) \right]\,,
\end{equation}
where $g_\alpha=G_Q\sum_n T_n$ is the conductance of the respective
tunnel junction junction, for which we have assumed that all $T_n\ll
1$ and $g_\alpha\gg G_Q$ to avoid Coulomb blockade. The Green's
function of the central node is determined by matrix current
conservation, reading $ \sum_{\alpha=1}^K \check I_\alpha = [
\sum_{\alpha=1}^K g_\alpha \check G_\alpha , \check G_c ]/2=0$.
Employing the normalization condition $\check G_c^2=1$, the solution
is
\begin{equation}
 \check G_c(\vec\chi) = \frac{\sum_{\alpha=1}^K g_\alpha \check G_\alpha(\chi_\alpha)}{
 \sqrt{\sum\nolimits_{\alpha,\beta=1}^K g_\alpha g_\beta
 \left\{\check G_\alpha(\chi_\alpha),\check G_\beta(\chi_\beta)\right\}}}\,.
\end{equation}
To find the cumulant-generating function (CGF) $S(\vec\chi)$ we
integrate the equations $\partial S(\vec\chi)/\partial
\chi_\alpha = (-it_0/4e^2) \int dE \mbox{Tr}\check\tau_K\check
I_{\alpha}(\vec\chi)$ \cite{nazarov:02-multi}.  We obtain
\begin{equation}
 \label{eq:cgf-general-multi}
 S(\vec\chi)
 = \frac{t_0}{2e^2}\int dE \mbox{Tr} 
 \sqrt{\sum\nolimits_{\alpha,\beta=1}^M g_\alpha g_\beta
 \left\{\check G_\alpha(\chi_\alpha),\check G_\beta(\chi_\beta)\right\}}\,.
\end{equation}
This is the general result for an M-terminal geometry in which 
all terminals are tunnel-coupled to a common node. It is valid for
arbitrary combinations of normal metal and superconductor, fully
accounting for the proximity effect. Note, that we have dropped the
normalization of $S(\vec\chi)$ to write the expression more compact.

\subsubsection{Normal metals}
If all terminals are normal metals, the matrices in
Eq.~(\ref{eq:cgf-general-multi}) are all diagonal and trace is trivial. 
We obtain
\begin{equation}
  \label{eq:cgf-multi-normal} 
  S(\vec\chi)= \frac{t_0}{ 2 e^2 } \int dE
  \sqrt{g_\Sigma^2 + \sum_{\alpha \neq \beta} g_\alpha g_\beta
       f_\alpha(E)(1-f_\beta(E))\left(e^{i(\chi_\alpha-\chi_\beta)}-1\right)}
\end{equation}
where $f_\alpha$ is the occupation function of terminal $\alpha$.
Here, we introduced the abbreviation
$g_{\Sigma}=\sum_{\alpha=1}^Ng_\alpha$ for the sum of all
conductances. We note, that the statistics is essentially
non-Poissonian, despite the fact the we are considering tunnel
junctions. 

We now restrict us to two terminals (in which case we have to
consider only one counting field $\chi=\chi_1-\chi_2$). For zero
temperature and voltage bias $V$ the CGF reads then
\begin{equation}
  \label{eq:cgf-double-tunnel-general}
  S(\chi)=\frac{t_0 V}{2e}\sqrt{g_\Sigma^2+ 4 g_1g_2(e^{i\chi}-1)},
\end{equation}
the result for a double tunnel junction first obtained by
de Jong \cite{dejong:96} using a master equation approach. 
We obtain as limiting cases for an asymmetric junction (either $g_1\ll
g_2$ or $g_1\gg g_2$) Poisson statistics $S(\chi)=(t_0V
g_1g_2/(g_1+g_2))(\exp(i\chi)-1)$. 

Next we consider a three terminal structure, which is voltage biased
such that the mean current $\bar{I}_3$ in lead $3$ vanishes (voltage
probe) and a transport current $\bar I=g_1g_2/(g_1+g_2)V$ flows
between terminals 1 and 2. The CGF is \cite{boerlindiplom}
\begin{eqnarray}
  \label{eq:ntgtn_s} 
  S(\vec\chi) & = & \frac{t_0 |V|}{2e}\left( 
    g_2 \sqrt{g_\Sigma^2+ 4g_3g_1(e^{-i\chi_1}-1) +
      4g_1g_2(e^{i\chi_2-i\chi_1}-1)}\right.\nonumber\\
  &&\left. +g_1\sqrt{g_\Sigma^2+ 4g_3g_2
   (e^{i\chi_2}-1) + 4g_1g_2(e^{i\chi_2-i\chi_1}-1)}\right)\,.
\end{eqnarray}
It is interesting to note that the presence of the voltage probe makes
the CGF asymmetric under the transformation $g_1\leftrightarrow g_2$,
whereas the current is symmetric.  In certain limits in which the
square roots in Eq.~\ref{eq:ntgtn_s} can be expanded one is able to
find the counting statistics. E.~g in the strong-coupling limit $g_3\gg
(g_1+g_2)$ we find
\begin{equation}
  S(\vec\chi) = \bar{N}\left[ e^{-i\chi_1}
    + e^{i\chi_2}-2\right] \label{eq:ntgtn_s_lim1}\,.
\end{equation}
The CGF is simply the sum of two Poisson distribution, demonstrating
drastically the effect of the voltage probe. It completely suppresses
the correlation between electrons entering and leaving the central node.

Another interesting geometry is a beam splitter configuration, in
which a voltage bias is applied between one terminal and the other
two. We find
\begin{equation}
  S^N(\chi_1,\chi_2)=\frac{t_0|V|}{2e}
  \sqrt{g_\Sigma^2+g_1g_3\left(e^{i\chi_1}-1\right) 
    + g_3g_2\left(e^{i\chi_2}-1\right)}\,.
  \label{eq:cgf-splitter-normal}
\end{equation}
In the limit that $g_1+g_2$ and $g_3$ are very different, we can
expand the CGF and find for the CGF
$S(\chi)=N_1e^{i\chi_1}+N_2e^{i\chi}$, i.~e., the tunneling processes
into the two terminals are uncorrelated. The corresponding
probability distribution is simply the product of two Poisson
distributions. 

\subsubsection{SN-contact}

We now consider the case of a double tunnel junction, in which one of
the terminals is superconducting. From the general result
(\ref{eq:cgf-general-multi}) and (\ref{eq:sneigenvalues}) we find
after some algebra
\begin{equation}
  \label{eq:cgf-double-tunnel-andreev}
  S(\chi)=\frac{t_0|V|}{e\sqrt 2} 
  \sqrt{g_1^2+g_2^2+\sqrt{\left(g_1^2+g_2^2\right)^2
      +4g_1^2g_2^2(e^{i2\chi}-1)}}\,.
\end{equation}
Remarkably, the statistics is fundamentally different from the
corresponding normal case (\ref{eq:cgf-double-tunnel-general}). Still,
the elementary events are transfers of pairs of electrons, which,
however are correlated in a more complicated way than normal
electrons. If the junction is very asymmetric, the FCS reduces to
Poissonian transfer of electron pairs. This is similar to the effect
of decoherence between electrons and holes for energies of the order
of the Thouless energy \cite{samuelsson:03-counting}.

For the beam splitter configuration we are also able to find the FCS
analytically. The CGF is \cite{boerlin:02}
\begin{eqnarray}
  \label{eq:cgf-splitter-andreev}
 \lefteqn{S(\chi_1,\chi_2)=
   \frac{Vt_0}{\sqrt{2}e}\times}\\\nonumber
 &&
 \sqrt{g_S^2+\sqrt{g_S^4+
 4g_3^2g_1^2(e^{i2\chi_1}-1)+
 4g_3^2g_2^2(e^{i2\chi_2}-1)+
 8g_3^2g_1g_2(e^{i(\chi_1+\chi_2)}-1)}},
\end{eqnarray}
where we abbreviated $g_S^2=g_3^2+(g_1+g_2)^2$. From this result we
see that the elementary processes are now double charge transfers to
either terminal of a splitting of a Cooper pair among the two
terminal. It is interesting to note, that, if we assume that $g_1+g_2$
and $g_3$ are very different (but $g_1\approx g_2$), we obtain
non-separable statistics
\begin{equation}
  \label{eq:cgf-splitter-andreev-limit}
  S(\chi)=N_{11} e^{i2\chi_1}+N_{22} e^{i2\chi_2}+N_{12} e^{i(\chi_1+\chi_2)}\,.
\end{equation}
This expression can not be written as a sum of two independent terms.
Furthermore, the last term is positive, which implies that current
crosscorrelation $S_{12}=-(2e^2/t_0)
(\partial^2/\partial\chi_1\partial\chi_2)
S(\chi_1,\chi_2)|_{\chi_1,\chi_2\to 0}$ are \textit{positive}.
Eq.~(\ref{eq:cgf-splitter-andreev-limit}) provides a simple
explanation for this surprising effect: it is a consequence of
independence of the different events, contributing to the current.
This result, in fact, holds for a large class of superconducting beam
splitters
\cite{belzig:03-incoherent,samuelsson:02-cavity,samuelsson:02-crosscorrelation,taddei:02-ferro}.

\section{Conclusion}
\label{sec:sum}

We have tried to give a pedagogical introduction to the field of
counting statistics. Many technical details were left out, but we have
tried to cover the essence of the derivation and concentrated on
looking at concrete examples. For a more thorough study we recommend
the recent book \textit{Quantum Noise in Mesoscopic Physics}
\cite{nazarov:03-book} or the original literature.  While a number of
aspects have already been explored, many open questions remain, e.~g.,
experimental strategies to measure FCS, strongly interacting systems,
or spin-dependent problems. For the future, we expect even more
activity in the field and, consequently, even more interesting results
will emerge.

\subsubsection{Acknowledgement}
The ideas presented here are the results of numerous discussions with
many people. In particular I would like to mention D. Bagrets, C.
Bruder, J.~C.  Cuevas, Yu. V. Nazarov, and P. Samuelsson. This work
was supported by the Swiss NSF and the NCCR Nanoscience.

\section{Appendix}

We summarize here the matrix-Green's function for superconducting and
normal contact, as they were used in the text. The time-dependent
Green's functions are expressed by their Fourier transforms $\check
G_0(t-t^\prime) = \int (dE/2\pi)$ $e^{-iE(t-t^\prime)}$$\check G_0(E)$.  The
energy-dependent Green's functions in the Keldysh$\times$Nambu-space
have the form
\begin{equation}
  \label{eq:reservoir}
  \check G(E)= \left( \begin{array}[c]{cc}
      (\bar A - \bar R) \bar f + \bar R & (\bar A - \bar R) \bar f \\
      (\bar A - \bar R) (1 - \bar f) & (\bar R - \bar A) \bar f + \bar A
    \end{array}\right)\,,
\end{equation}
where  the advanced, retarded and occupation Nambu matrices are
\begin{equation}
  \label{eq:nambu-matrices}
  \bar A(\bar R)= 
  \left( \begin{array}[c]{cc} 
      g_{A(R)} & f_{A(R)} \\
      f_{A(R)} & -g_{A(R)} \\
    \end{array}\right)
  \quad,\quad
  \bar f(E)=
  \left(\begin{array}[c]{cc} 
      f(E) & 0 \\
      0 & f(-E)
     \end{array}\right)\,.
\end{equation}
The phase $\varphi$ of the superconducting order parameter as well as the
electrical potential $eV$ enter via the gauge transformation $\check
G(t,t^\prime)=\check U(t) \check G_0(t-t^\prime) \check
U^\dagger(t^\prime)$. Here $\check U(t)=\exp\left[i\phi(t)\bar\tau_3/2\right]$,
where $\phi(t)=\varphi+eVt$.

In the calculation of the FCS of contacts between normal metals and
superconductors we frequently need the eigenvalues of anticommutators
of two Green's functions. For two normal metals $\{\check
G_{N1}(\chi),\check G_{N2}\}/2$ is diagonal and the eigenvalue is
\begin{equation}
  \label{eq:normaleigenvalues}
  \left[1+2f_{1}(E)\left(1-f_{2}(E)\right)\left(e^{i\chi}-1\right)
    +2f_{2}(E)\left(1-f_{1}(E)\right)\left(e^{-i\chi}-1\right)\right],
\end{equation}
for the electron block and the same expression with $E\to -E$ for the
'hole'-block in Nambu space. 

In the case of Andreev reflection, i.~e. for $eV,k_BT_{el}\ll\Delta$,
we find for $\{\check G_{N}(\chi),\check G_{S}\}/2$ the two eigenvalues
\begin{equation}
  \label{eq:sneigenvalues}
  \pm \sqrt{f_N(E)f_N(-E)\left(1-e^{i2\chi}\right)+
      (1-f_N(E))(1-f_N(-E))\left(1-e^{-i2\chi}\right)}.
\end{equation}

\end{document}